\documentclass[letterpaper, 10pt, conference]{ieeeconf}
\IEEEoverridecommandlockouts        
\usepackage{hyperref}

\title{\LARGE \bf Network Topology Change Identification from Inverse Covariance Matrices}

\title{\LARGE \bf Identifying Edge Changes in Networks from Inverse Covariance Matrices}

\title{\LARGE \bf Edge Changes Identification in Infrastructure Networks: A Sparse Total Least Squares Approach}

\title{\LARGE \bf Resilient Infrastructure Network: Sparse Edge Change Identification via $\ell_1$-Regularized Least Squares}

\author{Rajasekhar Anguluri
	\thanks{The author works with the Department of Computer Science and Electrical Engineering, University of Maryland, Baltimore County, MD 85281, USA (e-mail: \href{rajangul@umbc.edu}{rajangul@umbc.edu}). The author dearly thanks ChatGPT for enhancing the clarity and conciseness of the sentences.}}

\usepackage{graphics,color}
\usepackage[outdir=./]{epstopdf}
\usepackage{amsmath}
\usepackage{amssymb}
\usepackage{mathrsfs}
\usepackage{subfigure}
\usepackage[font=small,skip=5pt]{caption}
\usepackage{url}
\usepackage{float}
\usepackage{dsfont}
\usepackage{booktabs}
\usepackage{array}
\usepackage[table]{xcolor}
\let\labelindent\relax
\usepackage{enumitem}
\usepackage{optidef} 
\usepackage{float}

\usepackage{bbold} 

\newtheorem{theorem}{\bf \emph{Theorem}}[section]
\newtheorem{lemma}[theorem]{\bf \emph{Lemma}}

\newtheorem{proposition}[theorem]{Proposition}

\newtheorem{remark}{Remark}
\newtheorem{assumption}[theorem]{Assumption}
\newtheorem{example}{Example}

\newcommand{\bigzero}{\mbox{\normalfont\Large 0}}
\newcommand{\bigtimes}{\mbox{\normalfont\Large $\times$}}

\newcommand{\transpose}{\mathsf{T}} 

\newcommand*{\QEDB}{\hfill\ensuremath{\blacksquare}}
\newcommand*{\QEDW}{\hfill\ensuremath{\square}}

\newcommand{\what}{\widehat}

\newcommand{\kvec}{\mathrm{Vec}}
\newcommand{\kvech}{\mathrm{Vech}}

\newcommand{\mbf}{\mathbf}

\newcommand\oprocendsymbol{\hbox{$\square$}}
\newcommand\oprocend{\relax\ifmmode\else\unskip\hfill\fi\oprocendsymbol}

\usepackage{algorithmic}
\usepackage[linesnumbered,ruled,vlined]{algorithm2e}

\usepackage[linesnumbered,ruled,vlined]{algorithm2e}

\usepackage{tabstackengine}
\setstackEOL{\cr}

\begin{document}
	\maketitle
	
	\thispagestyle{empty} \pagestyle{empty}
	
\begin{abstract} 
\looseness=-1 
Adversarial actions and a rapid climate change are disrupting operations of infrastructure networks (e.g., energy, water, and transportation systems). Unaddressed disruptions lead to system-wide shutdowns, emphasizing the need for quick and robust identification methods. One significant disruption arises from edge changes (addition or deletion) in networks. 

We present an $\ell_1$-norm regularized least-squares framework to identify multiple but sparse edge changes using noisy data. We focus only on networks that obey equilibrium equations, as commonly observed in the above sectors. The presence or lack of edges in these networks is captured by the sparsity pattern of the weighted, symmetric Laplacian matrix, while noisy data are node injections and potentials. Our proposed framework systematically leverages the inherent structure within the Laplacian matrix, effectively avoiding overparameterization. 

We demonstrate the robustness and efficacy of the proposed approach through a series of representative examples, with a primary emphasis on power networks.

	\end{abstract}

	\section{Introduction}

Critical infrastructure network systems are those that are important to our societies \footnote{{United States of Department of Homeland Security lists sixteen sectors, including energy, water, manufacturing, as critical infrastructure systems \cite{cisa_critical_infrastructure}. Other countries might have an equivalent or even expanded list.}}. But for us, any such system is a graph having nodes and edges linking those nodes. At the basic level, many infrastructure networks satisfy conservation laws\footnote{Our usage of conservation laws is morally correct but is not rigorous as perceived by a theoretical physicist or a pure mathematician.} (a more appropriate word is equilibrium equations). These equations state: flows entering a network are neither created nor destroyed. Flows may be power, traffic, or fluid. The laws we study apply in the steady-state regime and are linear (governed by algebraic relations between edge flows and flows injected by the nodes) {\color{black} \cite{strang1988framework}}.

Identifying edge changes---additions, removals, or both---in networks means pinpointing node indices of the changed edges from noisy measurements. For finite-dimensional networks obeying equilibrium equations, these indices can be deduced from the sparsity pattern (zero and non-terms) of the symmetric Laplacian matrix. Specifically, the indices of the nodes of a removed edge is indicated by an off-diagonal element in the Laplacian that changes from non-zero to zero after the removal.
A related but very different problem that does not concern is the topology estimation: estimate edges present between all pairs of nodes (see \cite{rayas2022learning, seccamonte2023inference, deka2023learning}).

We restrict our attention to the undirected networks. Thus, the Laplacian is a symmetric, weighted matrix. Suppose that a disruptive event happens and certain edges are changed in an otherwise known reference network (we chiefly focus on edge removal exclusively and briefly comment on additions). Then, our goal is to identify those edge changes using noisy measurements of node potentials and injected flows. 

\looseness=-1 A naive approach involves estimating the entire network and comparing it with a reference, but this is inefficient when changes are sparse compared to the total number of edges. A more effective approach is to estimate these sparse edge changes directly. To achieve this, we develop a sparsity-based estimation framework, with the following contributions:


\begin{enumerate}
    \item Using the equilibrium equations, we develop a sparse error-in-variable type linear regression model in which the unknown sparse regression vector contains information about the edge changes. We take complete account of the structure implicit in the Laplacian matrix, thereby avoiding over-parameterization of the unknown vector.
    \item We develop two estimators to estimate the sparse 
    vector. The first is the $\ell_1$-norm regularized total least squares (TLS) estimator. To overcome the difficulties inherent in implementing TLS type estimators, in Theorem \ref{eq: main theorem}, we develop an equivalent reformulation, which allows for fast heuristics like proximal-gradient in \cite{arablouei2017fast}. 
    \item The second is the $\ell_1$-norm regularized least squares (the LASSO estimator). This estimator is applicable when the error in variables is not high (which happens when the noise in injected flows is not too large). 
\end{enumerate}

We validate LASSO's performance on a synthetic network and IEEE benchmark power systems. Despite our preliminary theoretical characterization of the sparse TLS estimator, we did not validate its performance empirically. This is due to the absence of standard solvers for implementing the TLS estimator, unlike the readily available ones for the LASSO estimator. Nonetheless, we identify this as an area for future exploration and development.

\textit{Related Research}: 
The authors in \cite{segarra2017network} use the eigenspectrum of the Laplacian matrix to discern topological variations in the network. In \cite{deka2020graphical, anguluri2024networks}, the authors provide an analytical characterization of the Laplacian matrix under edge additions and removals. Instead, \cite{rayas2023differential} uses sparse covariance matrix techniques to learn edge changes. Hypothesis testing-based identification methods are considered in \cite{bawa2022verification, bolognani2018grid}. Finally, edge identification in linear dynamical networks is considered in \cite{hao2021discernibility, RD-JAD-SR:15, parlangeli2021detection}. Studies in references \cite{ardakanian2019identification, brouillon2021bayesian} are close to our approach. They suggest using LASSO and sparse TLS-type estimation for the topology estimation, not for the edge change identification problem.

\section{Model and Definitions}\label{sec: prelims}

\subsection{Steady-state Model for Infrastructure Networks}
We abstract any infrastructure network as a graph, having $n$ nodes connected (or not) by $m$ edges. The graph has no multiple edges between nodes and no self-loops. Specify an orientation for each edge; that is, one node is the ``starting" node, and the other is the ``end" node. The incidence matrix $A\in \{-1,0,1\}^{m\times n}$ indicates which nodes belong to which edge.
Suppose the $i$-th edge's starting node is $j$ and the ending node is $k$. Then the $i$-th row of $A$ has $-1$ in column $j$ and $+1$ in column $k$. Thus, $A\mathds{1}_{n}=0$. 





Associate numbers $u_1,u_2,\ldots,u_{n}$ (called potentials) for nodes, numbers $w_1,w_2,\ldots,w_m$ (called flows) and positive weights $c_1,c_2,\ldots,c_m$ for edges in the graph. The potentials could represent the heights of nodes (in geodetic leveling networks), the pressures (in hydraulic networks), the voltages (in electrical networks), or the mass positions (in spring-mass networks). The flows could represent branch currents in electrical networks or the flow of commodities in hydraulic, traffic, or manufacturing networks. The inverse of the weights measures the resistance offered by edges to the flows (for e.g., in electrical networks, the weights are conductance). 

Let $u=(u_1,u_2,\ldots,u_{n})'$ be the vector of node potentials; $e=(e_1,e_2,\ldots,e_m)'$ the vector of edge potentials (potential difference on edges); $w=(w_1,w_2,\ldots,w_m)'$ be the vector of flows going through the edges; and $f=(f_1,f_2,\ldots,f_n)$ be the vector of injected flows. Define the diagonal matrix $C\in \mathbb{R}^{m\times m}$ whose diagonal entries specify weights given to the edges. Then, several infrastructure networks satisfy: 
\begin{itemize}
    \item \emph{Kirchhoff's law for edge flows:} $A^\transpose w=f$.
    \item \emph{Kirchhoff's law for potentials:} $e=-Au$.
    \item \emph {Constitutive relations (Ohm's or Hooke's law):} $w=Ce$.
\end{itemize}

These equations hold for the vectors $w$, $u$, and $f$ at any time instant. Let these vectors to be time-varying and combine the laws into one equilibrium (or balanced) equation: 
\begin{align}\label{eq: Laplacian equation}
    f(t)=A^\transpose C Au(t)=Lu(t).
\end{align}
Here, $L=A^\transpose C A$ is the $n\times n$ weighted symmetric Laplacian matrix. The equation in \eqref{eq: Laplacian equation} holds for complex-valued vectors and matrices (for e.g., AC power networks).  

The incidence matrix $A$ tells us edges incident to the nodes and the diagonal $C$ tells us the weights associated with the edges in the graph. By construction, the Laplacian $L$ specifies both the incident edges and the weights linked to these edges.
Thus, the sparsity pattern of \( L \) (indicating the positions of zero and non-zero entries) directly reveals the presence or absence of edges between nodes. Indeed, let $l_{ij}$ be the weight of the edge linking nodes $i$ and $j$ (this weight can be read off from the edge weight diagonal matrix $C$). Then,
 \begin{align}\label{eq: edge assignment}
L_{i, j}:= \begin{cases}-l_{i, j} & \text { if } i \neq j \\ \sum_{k \neq i} l_{i, k} & \text { if } i=j\end{cases}.  
 \end{align}
  If the nodes $i$ and $j$ have no edge, then $l_{i, j}=0$. The sparsity pattern (indices of zero and non-zero entries) of $L$ gives the edges in the network and calls this pattern the topology.

\subsection{Edge Changes Identification Problem}
The edge changes identification problem differs from the topology identification, focusing on determining the presence or absence of edges in the operating network compared to a reference. This reference network is a predetermined configuration known to the practitioner or the network configuration from the past, perhaps a few hours or days ago. We focus on the edge removal scenario because this is the most common concern in infrastructure networks. Examples include: line loss in power networks or leaky pipes in hydraulic networks. However, it is easy to extend our results to the edge addition and the mixed case, which involves additions and removals. 

\begin{figure}[t]
		\centering
		\includegraphics[width=0.95\linewidth]{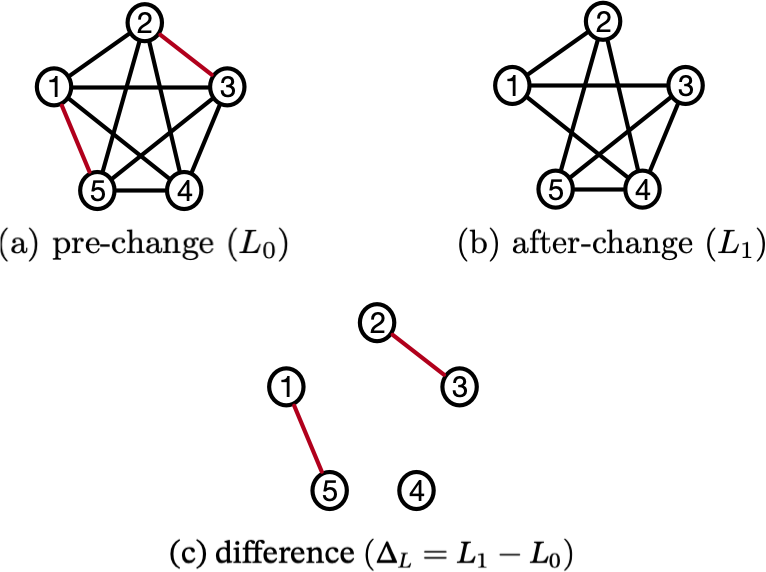}
		\caption {\small {\bf problem statement (illustrative)}: The graph underlying the Laplacian ${L}_0$ before the change with the network system is in (a). The Laplacian after edge changes is ${L}_1$, and is in (b). We want to directly estimate the graph underlying $\mathbf{\Delta}_{\mathbb{L}}$ in (c) (equivalently its sparsity pattern) using node potentials and injected flows.} \label{fig: problem statement}
	\end{figure}

Let $L_0$ and $L_1$ be Laplacians of the infrastructure network before and after a change\footnote{We assume that there is an inbuilt detection mechanism that alarms us if there is an edge change or (are changes) in the network, but what is unknown a priori is how many or which edges were changed.}. Let $\Delta L=L_1-L_0$. This sparsity pattern of $\Delta_L$ denotes edges removed after the change (see \ref{fig: problem statement} for an illustration and \cite{anguluri2024networks} for an algebraic characterization of $\Delta_L$). After the change, the balanced equation obeys
\begin{align}\label{eq: Laplacian equation after change}
    f(t)=L_1u(t)=(L_0+\Delta L)u(t).
\end{align}

For any $t$, let $\tilde{u}(t)=u(t)-\Delta u(t)$ and  $\tilde{f}(t)=f(t)-\Delta f(t)$ be the noisy measurements of the potentials $u(t)$ and injected flows $f(t)$ obeying the model in \eqref{eq: Laplacian equation after change}. Here, $\Delta u(t)$ and $\Delta f(t)$ are measurement errors. Use of the negative symbols in front of the errors will be clear in Section \ref{sec: main results}.
\begin{assumption}\label{assmp: problem setup}
    The pre-change Laplacian matrix $L_0$ and the noisy data after the change $\{\tilde{u}(t),\tilde{f}(t)\}$ are known. 
\end{assumption} 

Assuming \ref{assmp: problem setup}, the edge changes identification problem is to estimate the true sparsity pattern of $\Delta L$ using $\{\tilde{u}(t),\tilde{f}(t)\}$ collected over a finite time horizon after the change. Fig. \ref{fig: problem statement} states the problem pictorially for edges removed or missing from the reference network with Laplacian $L_0$.

\setlength{\textfloatsep}{2pt plus 1.0pt minus 2.0pt}

\section{A Sparse Linear Model Formulation for Edge Changes Identification}\label{sec: main results}
This section introduces a sparse total least squares (hereafter, sparse TLS) framework to estimate the sparsity pattern of $\Delta L$. We take account of the structure implicit in the Laplacian, avoiding over-parameterization in the TLS framework.


Let $t=1,\ldots,T$, and define the matrices of node potentials and injected flows after the change: 
\begin{align}
    \begin{split}
        U_T&=[u(1),u(2),\ldots,u(T)]\in \mathbb{R}^{n\times T}\\
        F_T&=[f(1),f(2),\ldots,f(T)]\in \mathbb{R}^{n\times T}.\\
    \end{split}
\end{align}
Then, the equilibrium equation in \eqref{eq: Laplacian equation after change} over the discrete time horizon can be succinctly expressed as 
\begin{align}\label{eq: matrix equilibrium}
    F_T=(L_0+\Delta L)U_T
\end{align}
Similar to the matrices $U_T$ and $F_T$ defined earlier, construct the $n\times T$ matrices $\tilde{U}_T$, $\tilde{F}_T$, $\Delta U_T$, and $\Delta F_T$ using the vectors $\tilde{u}(t)$, $\tilde{f}(t)$, $\Delta u(t)$, and $\Delta f(t)$. By definition $\tilde{F}_T=F_T-\Delta F_T$ and $\tilde{U}_T=U_T-\Delta U_T$. After some manipulations \eqref{eq: matrix equilibrium} becomes 
\begin{align}\label{eq: EIV-model}
    \tilde{F}_T+\Delta F_T=(L_0+\Delta L)(\tilde{U}_T+\Delta U_T), 
\end{align}
which is the so-called error-in-variable (EIV) model. 
In this model, $\tilde{F}_T$, $\tilde{U}_T$, and $L_0$ are known while the data errors $\Delta U_T$ and $\Delta F_T$, the difference matrix $\Delta L$ are unknown. 

\begin{assumption}\label{assump: sparse changes}
The number of edge changes are sparse; that is, $\|\Delta L\|_0\ll n^2$, where $\|\cdot\|_0$ is the counting norm.
\end{assumption}

Assumption \ref{assump: sparse changes} is reasonable for infrastructure systems as these are built to withstand extreme conditions. However, due to adversaries or unexpected operating conditions, a few edges might undergo disruptions.    

Under Assumptions in \ref{assmp: problem setup} and \ref{assump: sparse changes}, a straightforward way to estimate $\Delta_L$ in the presence of data errors is to solve the optimization problem: 
\begin{equation}\label{eq: naive problem}
\begin{aligned}
& \operatorname*{arg\, min}_{\Delta L=\Delta L^\transpose,\, \Delta F_T,\,\Delta U_T}
& & \|\Delta F_T\|_F^2+\|\Delta U_T\|_F^2+\lambda\|\Delta L\|_0 \\
& \quad \quad\,\, \text{subject to}
& & \text{EIV model in } \eqref{eq: EIV-model}. 
\end{aligned}
\end{equation}
Here $\|A\|_F=\sqrt{\mathrm{tr}(AA^\transpose)}$ is the Frobienus norm and $\lambda >0$ is the regularization parameter. However, the problem in \eqref{eq: naive problem} is a non-convex, combinatorial optimization, which is hard to solve in practice. A computationally superior alternative is to consider a convex relaxation to \eqref{eq: naive problem} by replacing $\|\cdot\|_0$-norm with the $\ell_1$ norm; that is $\|\Delta_L\|_1=\sum_{i,j}|\Delta_L(i,j)|$. 


While the $\ell_1$-relaxed problem is computationally simpler to solve, there are still problems with the naive formulation in \eqref{eq: naive problem}. First, the optimization in \eqref{eq: naive problem} accounts for symmetry and sparsity of $\Delta_L$ but fails to account the structured sparsity of $\Delta_L$ (that is, zeros of $\Delta L$ correspond to presence or absence of edges common to $L_0$ and $L_1$). To systematically exploit these properties, we convert the matrix-valued EIV in \eqref{eq: EIV-model} to the matrix-vector form in which $\Delta L$ is a vector.





\subsection{Vectorization of the EIV model}\label{subsec: vec EIV}
To reduce the notation clutter, we rewrite $L_0+\Delta_L$ as $L_1$ in \eqref{eq: EIV-model}. Recall that $A\otimes B$ is the Kroneckor product of matrices $A$ and $B$. Applying $\kvec(\cdot)$ on both sides of \eqref{eq: EIV-model} gives
\begin{align}\label{Eq: full vectorization}
      \kvec(\tilde{F}_T+\Delta F_T)&=\kvec(L_1(\tilde{U}_T+\Delta U_T))\nonumber \\
    &=((\tilde{U}_T+\Delta U_T)^\transpose \otimes \mathds{I}_n)\kvec(L_1) \\
    &=[(\tilde{U}^\transpose_T\otimes \mathds{I}_n)+ ((\Delta U_T)^\transpose\otimes \mathds{I}_n)]\kvec(L_1).\nonumber 
\end{align}

The $n\times n$ symmetric matrix $L_1$ in \eqref{eq: edge assignment}, including diagonal, has $n(n+1)/2$ free parameters and $n(n-1)/2$ repeats. We assume these repeats are from the lower triangular portion of $L_1$. Thus, $\kvec(L_1)$ is $n^2$-dimensional vector with $n(n-1)/2$ repeats. Let $\kvech(L_1)$ be the $n(n+1)/2$-dimensional vector obtained by deleting the repeats in $\kvec(L_1)$. Formally,  depending on $n$, there exists a unique duplication matrix $D$ with $\kvec(L_1)=D\kvech(L_1)$ \cite{magnus1980elimination}. The elimination matrix $E$ reverses the duplication operation: $\kvech(L_1)=E\kvec(L_1)$. Algebraic properties of $D$ and $E$ are documented in \cite{magnus1980elimination}.

We express $\kvech(L_1))$ in $L_0$ and $\Delta L$ using the linearity of the vectorization operator. Thus, $\kvech(L_1)=\kvech(L_0)+\kvech(\Delta L)$. Further, owing to the linearity, the left-hand side of \eqref{Eq: full vectorization} can also be expanded as $\kvec(\tilde{F}_T+\Delta F_T)=\kvec(\tilde{F}_T)+\kvec(\Delta F_T)$. Putting the pieces together, write \eqref{Eq: full vectorization} as 
\begin{align*}
\kvec(\tilde{F}_T)+\kvec(\Delta F_T)=\nonumber\\
    &\hspace{-30.0mm}[\tilde{U}^\transpose_T\otimes \mathds{I}_n+ (\Delta U_T)^\transpose\otimes \mathds{I}_n]D[\kvech(L_0)+\kvech(\Delta L)]. 
\end{align*}
We drop the subscript $T$ and express the above equality using the compact notation (defined in the table below): 
\begin{align}\label{eq: compact EIV}
    {y}+\Delta y=({X}+\Delta X)(\beta_0+\beta), 
\end{align}
where ${X}$, ${y}$, and $\beta_0$ are known; $\Delta X$ and $\Delta y$ are unknown data errors; and $\beta$ is the unknown sparse vector. 
\begin{table}[h!]
\normalsize  
    \centering
     \caption{{\normalsize Summary of matrices and vector notations}}
    \begin{tabular}{|c|c|}
    \hline 
       Matrices & Vectors\\
         \hline 
        $\begin{aligned}
    {X}&=(\tilde{U}^\transpose_T\otimes \mathds{I}_n)D\\
    \Delta X &=((\Delta U_T)^\transpose\otimes \mathds{I}_n)D
\end{aligned}$ & $\begin{aligned}
    \beta_0&=\kvech(L_0)\\
    \beta & = \kvech(\Delta L)\\
    {y}&=\kvec(\tilde{F}_T)\\
    \Delta y&=\kvec(\Delta F_T)
\end{aligned}$\\
\hline 
    \end{tabular}
    \label{tab:my_label}
\end{table}

\vspace{-2.0mm}
In the passage preceding Section \ref{subsec: vec EIV}, we mentioned that $\Delta L$ is structured sparse: the non-zeros of $\Delta_L$ correspond to the edges either absent or present in the network before the change. To exploit this sparsity, without loss of generality partition\footnote{This can be done by renumbering the nodes such that $\beta_s$ contain non-zero terms in $L_0$. Alternatively, we can introduce a permutation matrix $\Gamma$ such that $\beta_0=\Gamma [\beta_{0,s}^\transpose\, 0^\transpose]^\transpose$.} $\beta_0$ as $\beta_0=[\beta_{0,s}^\transpose\, 0^\transpose]^\transpose$, where $\beta_s$ contains non-zero terms in the lower triangular part of $L_0$ (including diagonals). Then $\beta$ can be partitioned as $[\beta_{s}^\transpose\, 0^\transpose]^\transpose$. But unlike the dense\footnote{By construction, $\beta_{s,0}$ will not even have one zero entry; see Fig.~\ref{fig: synthetic sparsity}.} $\beta_{s,0}$, the vector $\beta_s$ is sparse. Thus, \eqref{eq: compact EIV} becomes 
\begin{align}
    {y}+\Delta y&=(\begin{bmatrix}
        X_s & X_{s'}
    \end{bmatrix}+\begin{bmatrix}
        \Delta X_s & \Delta X_{s'}
    \end{bmatrix})\begin{bmatrix}
        \beta_{s,0}+\beta_{s}\nonumber \\
        0
    \end{bmatrix}, \\
    &=(X_s+\Delta X_s)(\beta_{s,0}+\beta_{s})\label{eq: reduced EIV}
\end{align}

The model in \ref{eq: reduced EIV} is minimal in that there are no redundant parameters in $\beta_s$ or edges not present in the network. For example, in the IEEE 118 bus system, $n=118$; but there are only $358$ edges. Thus, dimension of $\beta_0$ is $n(n+1)/2=7021$. Instead, the dimension of $\beta_s$ (including diagonals) is $476$, which is small. (See Fig.~\ref{fig: synthetic sparsity} for a visual description). The savings in dimensionality become evident in large-scale systems with sparse connections and fewer edge changes.

\section{Solution Strategies}
The techniques in this section hold for models in \eqref{eq: compact EIV} and \eqref{eq: reduced EIV}. For ease of notation, we stick with the model in \eqref{eq: compact EIV}. 


Suppose that data errors $\Delta y$ and $\Delta X$ are non-negligible. Then, a sparse estimate of $\beta$ can be obtained by solving the sparse variant of the TLS problem:
\begin{align}
\textbf{(P.1)}\quad & \hat{\beta}=\operatorname*{arg\, min}_{\Delta y,\, \Delta X,\, \beta}
& & \|\Delta y\|_2^2+\|\Delta X\|_F^2+\lambda \|  \beta\|_1\nonumber \\
& \text{subject to}
 &&  \hspace{-2.0mm} y+\Delta y=(X+\Delta X)(\beta_0+\beta). \nonumber
\end{align}

The regularizer is $\| \beta\|_1=\sum_i|\beta_i|$. The changed edges can be identified from the support of $\hat{\beta}$ returned by solving (P.1). Note that (P.1) is not equivalent to the naive optimization in \eqref{eq: naive problem} because we accounted for the structured sparsity of $\Delta L$ in the former problem via the vector $\beta$ or $\beta_s$. 

The TLS problem without the $\ell_1$-norm regularizer is non-convex due to $\Delta X\beta_0$ in the equality constraint. Nonetheless, the unique solution is computed in a closed form using the singular value decomposition (SVD) of the matrix $[\Delta X\,\, \Delta y]$ \cite{markovsky2007overview}. 
Unfortunately, such SVD-type solutions do not exist in the presence of the $\ell_1$-regularizer term. Recently, \cite{zhu2011sparsity} suggested a reformulation for a problem akin to (P.1), allowing the use of greedy methods. We generalize \cite[Lemma 2.1]{zhu2011sparsity}.

\begin{theorem}\label{eq: main theorem}
The constrained optimization problem in \textbf{(P.1)} is equivalent to solving the unconstrained optimization problem involving only the variable $\beta$: 
\end{theorem}
\begin{align}\label{Eq: final problem}
    \hat{\beta}=\operatorname*{arg\,min}_{\beta} \frac{\|X(\beta_0+\beta)-y\|^2_2}{1+\|(\beta_0+\beta)\|_2^2}+\lambda \|\beta\|_1
\end{align}
\begin{proof}
    We follow the technique in \cite[Lemma 2.1]{zhu2011sparsity}. Let $v=\kvec([\Delta X\,\, \Delta y])$ and write the first two terms in the cost function of (P.1) as $\|\Delta y\|_2^2+\|\Delta X\|_F^2=[\Delta X\,\, \Delta y]_F^2=\|v\|_2^2$. Next, rewrite the constraint in (P.1):
    \begin{align*}
        y-X(\beta_0+\beta)&=\Delta X(\beta_0+\beta)-\Delta y\\
        &/\begin{bmatrix}
        \Delta X & \Delta y
        \end{bmatrix}\begin{bmatrix}
            (\beta_0+\beta)\\
            -1
        \end{bmatrix}. 
    \end{align*}
Using the crucial fact that $Ax=\kvec(Ax)=(x^\transpose \otimes I)\kvec(A)$ for some matrix $A$ and $x$, we have
    \begin{align}\label{eq: proof constraint}
      \hspace{-2.5mm}  y-X(\beta_0+\beta)&=\underbrace{([
            ((\beta_0+\beta))^\transpose\,\, -1
        ]\otimes I)}_{G(\beta)}v. 
    \end{align}

For any fixed $\beta$, the $\ell_1$-norm constrained can be dropped in (P.1), and the resulting optimization problem then becomes $\operatorname{min}_{v}\|v\|_2^2$ subject to the constraint in \eqref{eq: proof constraint}---the least norm problem. The solution to this problem is $v(\beta)=G(\beta)^\dagger [y-X(\beta_0+\beta)]$. Using the special structure of $G(\beta)$ we further simplify the pseudoinverse $G^\dagger(\beta)$ as
\begin{align*}
 G^\dagger(\beta)&= G^\transpose(\beta)[G(\beta)G^\transpose(\beta)]^{-1} \\
 &= G^\transpose(\beta)[(1+\|(\beta_0+\beta)\|_2^2)\otimes I]^{-1}\\
 &=G^\transpose(\beta)(1+\|(\beta_0+\beta)\|_2^2)^{-1}. 
\end{align*}
Thus, $v(\beta)=(1+\|(\beta_0+\beta)\|_2^2)^{-1}G^\transpose(\beta)[y-X(\beta_0+\beta)]$ for fixed $\beta$. Plugging this expression of $v(\beta)$ back in  $\|v\|_2^2$, and simplifying it, gives the equivalent form in \eqref{Eq: final problem}. 
\end{proof}

Compared to (P.1), the optimization in \eqref{Eq: final problem} is computationally simpler since it involves only one vector variable $\beta$. Nevertheless, the problem remains non-convex and necessitates heuristic methods for its resolution. We leave this task to future work as it falls outside the scope of this paper.



\subsection{$\ell_1$-regularized least squares: LASSO estimator} 
Suppose that the error $\Delta X$ is small compared to $\Delta y$, then we may ignore it from \eqref{eq: EIV-model} to get  
\begin{align}\label{eq: compact EIV2}
    {y}-X\beta_0={X}\beta-\Delta y. 
\end{align}
Since $X$ and $\beta_0$ are known, the left-hand side is completely determined. Thus, the model in \eqref{eq: compact EIV2} is a simple linear model with the additive noise term $\Delta y$. Then, the sparse estimate of $\beta$ can obtained by solving the $\ell_1$-regularized least squares problem (or familiarly called the LASSO estimator): 
\begin{align}
\textbf{(P.2)}\quad & \hat{\beta}=\operatorname*{arg\, min}_{\beta} \,\, 
\frac{1}{2}\|y-X\beta_0-X\beta\|_2^2+\lambda \|  \beta\|_1 \label{eq: LASSO estimator}
\end{align}

\vspace{-2.0mm}
Unlike the optimization problems we have seen thus far, (P.2) is a convex problem, and several numerical methods are proposed in the literature to compute $\hat{\beta}$ \cite{zhao2023survey}. The most reliable and fast method is the coordinate descent method (available in standard software packages such as MATLAB). 


\section{Simulations}

We analyze the performance of the LASSO estimator \eqref{eq: LASSO estimator} on a synthetic network and on three real-world power system networks (IEEE-57, IEEE-118, IEEE-145; the number here denotes the total nodes). We set $T=30$ (the measurement window) and generate error terms $\Delta u(t)$ and $\Delta f(t)$ in \eqref{eq: matrix equilibrium} using zero-mean Gaussian distribution with variance $0.1$. 

We rely on the indexes below to quantify the performance of the estimator. The symbol $\wedge$ is the $\operatorname{AND}$ operator. 
\begin{itemize}
    \item $\operatorname{TP}$: proportion of the removed edges correctly identified ($\widehat{{\beta}}(i)\ne 0\wedge {{\beta}}(i)\ne0$)
    \item $\operatorname{TN}$: proportion of the non-removed edges correctly identified ($\widehat{{\beta}}(i)= 0\wedge {{\beta}}(i)=0$)
    \item $\operatorname{FN}$: proportion of the removed edges incorrectly identified ($\widehat{{\beta}}(i)\ne 0\wedge {{\beta}}(i)=0$)
    \item $\operatorname{FP}$: proportion of the non-removed edges incorrectly identified ($\widehat{{\beta}}(i)= 0\wedge {{\beta}}(i)\ne 0$). 
\end{itemize}

\bigskip   
We report  $\operatorname{acc}=\operatorname{TP}+\operatorname{TN}/(\operatorname{TP}+\operatorname{TN}+\operatorname{FP}+\operatorname{FN})$ measure with $0\leq \text{acc} \leq 1$ to summarize the indexes. The higher the $\operatorname{acc}$ measure, the better the performance. We averaged our results over twenty independent runs (standard deviation is not reported since it is not informative). Our results suggest that for a suitable $\lambda>0$, sparsity patterns of $\hat{\beta}$ and $\beta$ match.

	\begin{figure}
		\centering
		\includegraphics[width=1.0\linewidth]{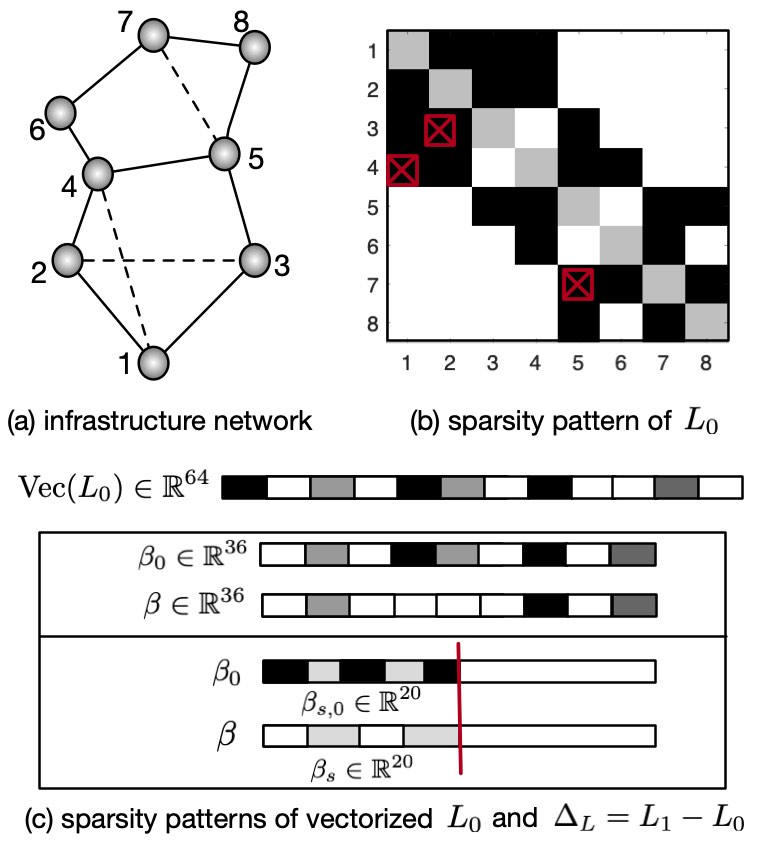}
		\caption {\small {\bf Visualizing synthetic network ($8$ nodes and $12$ edges)}: (a) The edges shown in dashed lines are removed after the change. (b) White spaces correspond to nodes with no edges; making red check boxes white (including a copy on the upper triangular portion) gives the sparsity pattern for $L_1$. Sparsity patterns in (c) are representative only and should not be confused with the no. of boxes in the vectors and the dimensions of $\beta$'s or $\operatorname{Vec}(L_0)$ denoting these vectors; see second paragraph in \ref{sec: synthetic network} for additional details.} \label{fig: synthetic sparsity}
	\end{figure} 

\begin{figure}[h!]
		\centering
		\includegraphics[width=0.9\linewidth]{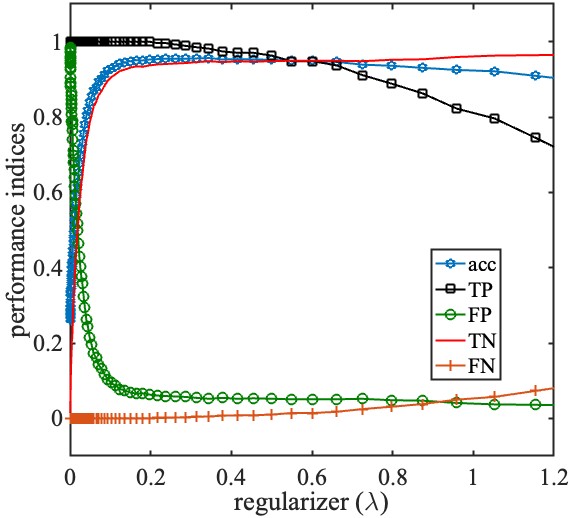}
		\caption {\small {LASSO performance on the synthetic network in Fig.~\ref{fig: synthetic sparsity}. The optimal $\lambda$ is for which acc, TP, and TN are closer to one and FP and FN are closer to zero}. Thus, the optimal range is $0.15\leq \lambda \leq 0.6$.} \label{fig: synthetic scores}
	\end{figure} 

\subsection{Synthetic network}\label{sec: synthetic network}
Consider the network shown in Fig.~\ref{fig: synthetic sparsity} (a), consisting of eight nodes and twelve edges---three removed after a change ($\{(2,3), (4,1), (7,5)\}$). Sparsity patterns of $L_0$ (pre-change) and $L_1$ (after-change) are in Fig.~\ref{fig: synthetic sparsity} (b). And Fig.~\ref{fig: synthetic sparsity} (c) shows $\operatorname{Vec}(L_0)$, its reduced vector $\beta_0$, and the difference vector $\beta$. The bottom box shows $\beta_0$ and $\beta$ after a suitable permutation of nodes; and the white space to the right of the vertical line are zeros. By construction, $\beta_{s,0}$ has twenty non-zero terms (twelve edges plus eight diagonals in $L_0$), and $\beta_s$ have six non-zeros (three terms below and on the diagonal of $\Delta L$); see Section \ref{subsec: vec EIV} for details. 

From Table \ref{table-sparse}, the computationally efficient model is \eqref{eq: reduced EIV} because the vector to be estimated has a lesser dimension. But we work with the model in \eqref{eq: compact EIV} for simplicity and solve (P.2) in \eqref{eq: LASSO estimator} using MATLAB's built-in command LASSO for $\widehat{\beta}$. Fig.\ref{fig: synthetic scores} shows performance indices as a function of $\lambda$. 

\begin{table}
\normalsize 
    \centering
    \caption{\label{table-sparse}{\normalsize Number of non-zeros in $\operatorname{Vec}(\Delta L)$ and $\beta$}}
    \begin{tabular}{|cccc|}
    \hline 
       vector  & dimension & $\#$ of non-zeros & model \\
    \hline 
       $\operatorname{Vec}(\Delta L)$  & 64 & 6 & Eq \eqref{Eq: full vectorization}\\
       $\beta$  & 36 & 6 & Eq \eqref{eq: compact EIV}\\
       $\beta_s$  & 20  & 6  & Eq \eqref{eq: reduced EIV}\\
       \hline 
    \end{tabular}
\end{table}

\subsection{Power system networks}
\begin{figure}[t]
		\centering
		\includegraphics[width=0.95\linewidth]{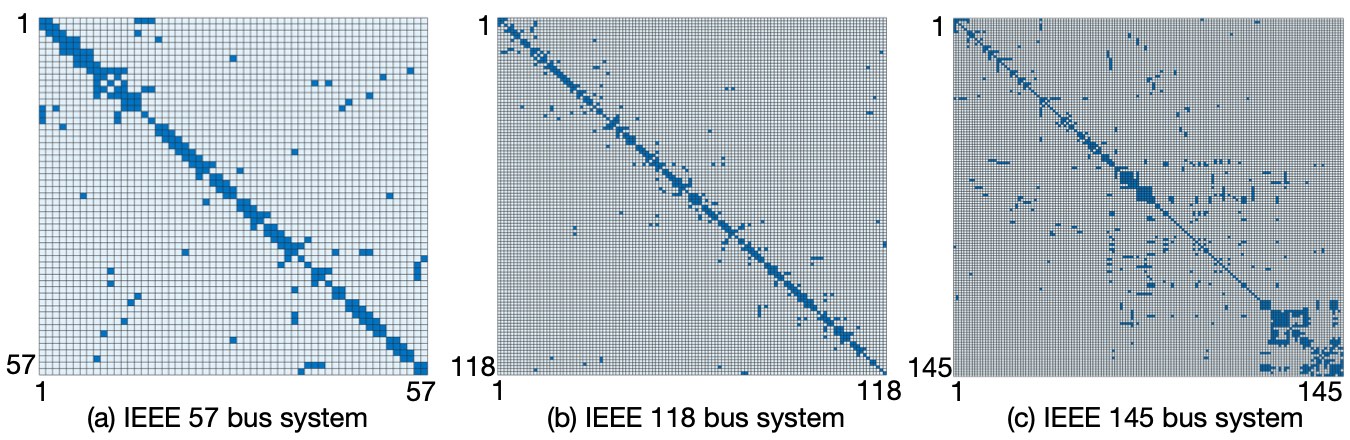}
		\caption {\small {\bf Sparsity patterns of power networks' Laplacian matrices $(L_0)$ before change}: The dimensions of matrices are $(57\times 57)$, $(118\times 118)$, and $(145\times 145)$, from left to right. The dark-colored boxes denote non-zeros, while the light-shaded boxes denote zeros.} \label{fig: power-sparsity}
	\end{figure}

\begin{figure}
		\centering
		\includegraphics[width=1.0\linewidth]{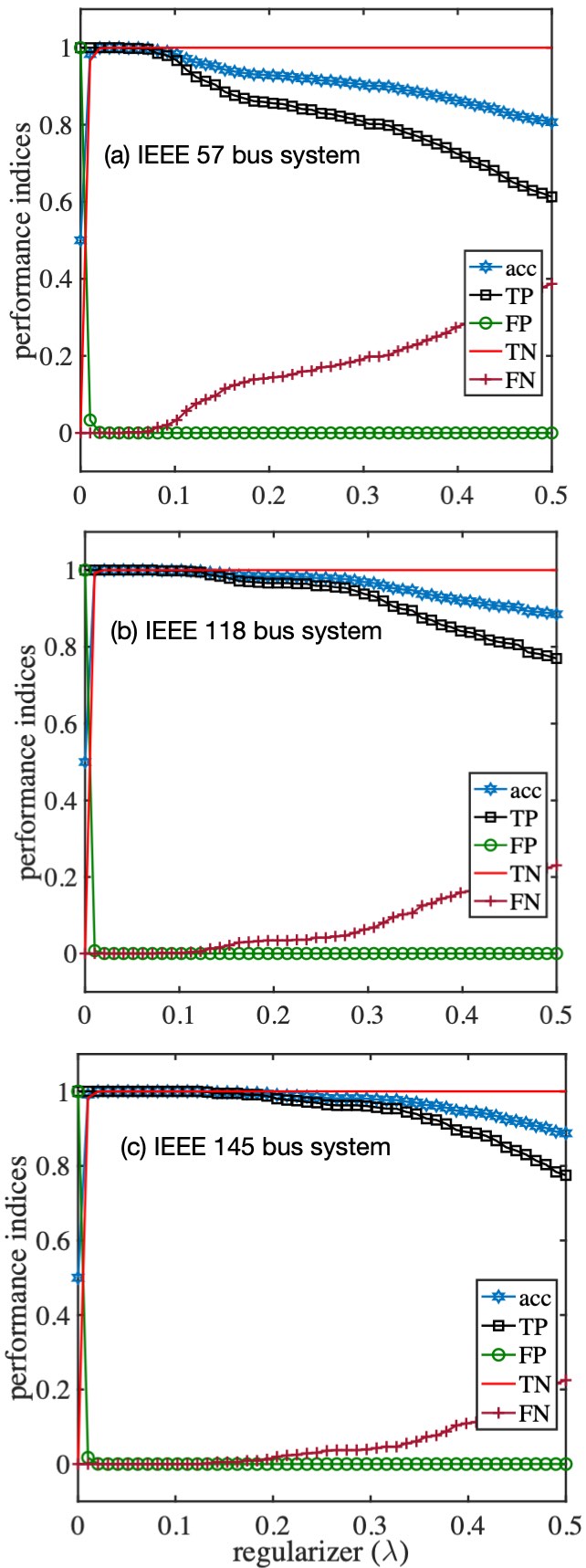}
		\caption {\small {LASSO performance on power system networks in Fig.~\ref{fig: power-sparsity}. The optimal range of $\lambda$ are the ones for which acc, TP, and TN are closer to one, and FP and FN are closer to zero}. Thus, the optimal range for all three systems is $0.1\leq \lambda \leq 1$.} \label{fig: power-results}
	\end{figure}

The sparsity patterns of Laplacian matrices $(L_0)$ of power system networks before edge change changes are shown in \ref{fig: power-sparsity}. We use MATPOWER (open-free software) to obtain $L_0$. For each network, we randomly remove ten edges and compute $L_1$ such that \eqref{eq: edge assignment} holds. Thus, true $\beta$ has at most $20$ zeros (ten from off-diagonals and another ten from diagonals of $\Delta L$).  Fig.\ref{fig: power-results} shows the performance of the LASSO estimator in \eqref{eq: LASSO estimator} subject to the model in \eqref{eq: compact EIV}. For all networks, the optimal range of $\lambda$ values (based on the acc measure) that result in better performance indices is between $0$ and $0.15$. 

\vspace{-1.0mm}

\section{Conclusions}
This work presented two $\ell_1$-norm regularized least-squares estimators to identify edge changes in critical infrastructure networks obeying equilibrium equations. The first one was a sparse total least squares (TLS) estimator, applicable when node potentials and injected flows were corrupted by noises. Theorem \ref{eq: main theorem} provided a simple reformulation for the sparse TLS estimator. The second one was the well-known LASSO estimator, applicable if the noise of the injected flow was not high (in magnitude). The LASSO estimator was convex and computationally efficient; instead, the TLS estimator was non-convex and needs heuristic methods, which will address in our future research. 

\bibliographystyle{unsrt}
\bibliography{BIB.bib}

\end{document}